\documentclass[a4paper,10pt]{article}

 \usepackage{epsf}
 \usepackage{graphicx}    % Include figure files

 \newcommand{\be}[1]{\begin{equation}\label{#1}}
 \newcommand{\ee}{\end{equation}}
 \newcommand{\bea}{\begin{eqnarray}}
 \newcommand{\eea}{\end{eqnarray}}

 \def\gsim{ \lower .75ex \hbox{$\sim$} \llap{\raise .27ex \hbox{$>$}} }
 \def\lsim{ \lower .75ex \hbox{$\sim$} \llap{\raise .27ex \hbox{$<$}} }

 \renewcommand{\markright}{\markright{\thepage}}

\begin{document}

 \begin{titlepage}

 \begin{flushright}
 arXiv:0802.4122
 \end{flushright}

 \vspace{5mm}

 \begin{center}

 {\Large \bf Growth Index of DGP Model and Current Growth Rate Data}

 \vspace{10mm}

 {\large Hao~Wei$^{\,}$\footnote{\,email address:
 haowei@mail.tsinghua.edu.cn}}

 \vspace{2mm}
 {\em Department of Physics and Tsinghua Center for Astrophysics,\\
 Tsinghua University, Beijing 100084, China}

 \end{center}

 \vspace{5mm}
 \begin{abstract}
Recently, some efforts focus on differentiating dark energy and
 modified gravity with the growth function $\delta(z)$. In the
 literature, it is useful to parameterize the growth rate
 $f\equiv d\ln\delta/d\ln a=\Omega_m^\gamma$ with the growth
 index $\gamma$. In this note, we consider the general DGP model
 with any $\Omega_k$. We confront the growth index of DGP model
 with currently available growth rate data and find that the DGP
 model is still consistent with it. This implies that more and
 better growth rate data are required to distinguish between
 dark energy and modified gravity.\\

 \end{abstract}

 \end{titlepage}

 \newpage

 \setcounter{page}{2}

%============================= section 1 ===================================
\section{Introduction}\label{sec1}
The current accelerated expansion of our
 universe~\cite{r1,r2,r3,r4,r5,r6,r7,r8,r9,r49} has been one of the
 most active fields in modern cosmology. There are very strong
 model-independent evidences~\cite{r10} (see also e.g.~\cite{r11})
 for the accelerated expansion. Many cosmological models have been
 proposed to interpret this mysterious phenomenon, see
 e.g.~\cite{r1} for comprehensive reviews.

In the flood of various cosmological models, one of the most
 important tasks is to discriminate between them. Recently,
 some efforts have been made. For instance, it is important to
 determine that the dark energy is cosmological constant or
 dynamical dark energy~\cite{r1}. Caldwell and Linder proposed
 a so-called $w-w^\prime$ analysis in~\cite{r12} to discriminate
 dark energy models, and then was extended in~\cite{r13,r14}.
 A recent review on $w-w^\prime$ analysis can be found
 in~\cite{r15}. Another tool to discriminate models is the
 statefinder diagnostic proposed by Starobinsky~{\it et al.}
 in~\cite{r16}. For a comprehensive list of relevant works on
 $w-w^\prime$ analysis and statefinder diagnostic, one can see
 e.g.~\cite{r17} and references therein.

Recently, some efforts to discriminate models focus on
 differentiating dark energy and modified gravity with the growth
 function $\delta(z)\equiv\delta\rho_m/\rho_m$ of the linear matter
 density contrast as a function of redshift $z$. By now, most of
 cosmological observations merely probe the expansion history of
 our universe~\cite{r1,r2,r3,r4,r5,r6,r7,r8,r9,r49}. As well-known,
 it is very easy to build models which share a same cosmic expansion
 history by means of reconstruction between models. Therefore, to
 distinguish various models, some independent and complementary
 probes are required. Recently, it is argued that the measurement
 of growth function $\delta(z)$ might be competent, see
 e.g.~\cite{r18,r19,r20,r21,r22,r23,r24,r25,r26,r27,r28,r29,r30,r31,r38}.
 If two models, especially dark energy and modified gravity models,
 share a same cosmic expansion history, they might have
 {\em different} growth history. Thus, they could be distinguished
 from each other.

One of the leading modified gravity models is the so-called
 Dvali-Gabadadze-Porrati~(DGP) braneworld model~\cite{r32,r33},
 which altering the Einstein-Hilbert action by a term arising from
 large extra dimensions. For a list of references on DGP model, see
 e.g.~\cite{r29,r34} and references therein. The first approach to
 study the growth function $\delta(z)$ of DGP model is numerical
 solution, see e.g.~\cite{r25,r27,r35}. The second approach is
 to parameterize the growth rate $f\equiv d\ln\delta/d\ln a$, where
 $a=(1+z)^{-1}$ is the scale factor of our universe.

For many years, it has been known that a good approximation to
 the growth rate $f$, within Einstein gravity, is given by
 \be{eq1}
 f\equiv\frac{d\ln\delta}{d\ln a}=\Omega_m^\gamma,
 \ee
 where $\gamma$ is the growth index, whereas $\Omega_m$ is the
 fractional energy density of matter. In the very beginning,
 Eq.~(\ref{eq1}) was introduced in~\cite{r36,r37}. There, $f$ was
 defined purely in terms of the present value, using the present
 matter density, not valid for arbitrary redshift. Also, it was
 not until~\cite{r21} that it was applied to anything beyond
 matter, curvature, and a cosmological constant. Finally, not
 until~\cite{r28} was it applied to gravity other than general
 relativity, and then in~\cite{r19} generalized to modified
 gravity, varying equation of state, and an integral relation for
 growth. This parameterized approach has been tested in some works
 recently, see e.g.~\cite{r18,r19,r20,r22,r28,r29,r30,r31,r38}.
 The theoretical value of $\gamma$ for $\Lambda$CDM model is
 $6/11\simeq 0.545$~\cite{r18,r19} whereas $\gamma\simeq 0.55$ for
 other parameterized dark energy models~\cite{r19}. The theoretical
 growth index of flat DGP model (whose $\Omega_k=0$ exactly) is
 $\gamma=11/16=0.6875$~\cite{r18} (In fact, $\gamma=11/16$ is
 a high redshift asymptotic value. $\gamma=0.68$ is favored for the
 fit to the whole growth history to the present, while another
 approximation is given in Eq.~(27) of~\cite{r18} for any redshift,
 with $\gamma=7/11$ in the asymptotic future). Therefore,
 it is possible to distinguish the dark energy model (including
 $\Lambda$CDM model) from the flat DGP model.

In this note, we consider the general DGP model with any $\Omega_k$.
 We confront the growth index of DGP model with currently available
 growth rate data and find that the DGP model is still consistent
 with it. On the other hand, it is shown that the $\Lambda$CDM
 model is also consistent with current growth rate data~\cite{r31}.
 This implies that more and better growth rate data are required to
 distinguish between dark energy and modified gravity models.

%============================= section 2 ===================================
\section{Growth index of DGP model}\label{sec2}
In the DGP model, the Friedmann equation is modified
 as~\cite{r33} (see also e.g.~\cite{r28,r29,r39})
 \be{eq2}
 H^2+\frac{k}{a^2}-\frac{1}{r_c}\sqrt{H^2+\frac{k}{a^2}}
 =\frac{8\pi G}{3}\rho_m,
 \ee
 where $H\equiv\dot{a}/a$ is the Hubble parameter; a dot denotes the
 derivative with respect to cosmic time $t$; the constant $r_c$ is
 the crossover scale; spatial curvature $k=0$, $k>0$ and $k<0$ for
 flat, closed and open universe, respectively. Here, we only consider
 the self-accelerating branch. It is easy to rewrite Eq.~(\ref{eq2})
 as an equivalent form~\cite{r33} (see also e.g.~\cite{r40})
 \be{eq3}
 H^2+\frac{k}{a^2}=\frac{8\pi G}{3}\left(\sqrt{\rho_m+\rho_{r_c}}
 +\sqrt{\rho_{r_c}}\right)^2,
 \ee
 where $\rho_{r_c}\equiv 3/(32\pi G r_c^2)$. One can alternatively
 recast Eq.~(\ref{eq2}) as the ``standard'' form
 \be{eq4}
 H^2+\frac{k}{a^2}=\frac{8\pi G}{3}\left(\rho_m+\rho_{de}\right),
 \ee
 where
 \be{eq5}
 \rho_{de}=\frac{3}{8\pi Gr_c}\sqrt{H^2+\frac{k}{a^2}}
 =2\sqrt{\rho_{r_c}}\left(\sqrt{\rho_m+\rho_{r_c}}
 +\sqrt{\rho_{r_c}}\right),
 \ee
 in which we have used Eq.~(\ref{eq3}) in the last equality. Here,
 $\rho_{de}$ can be regarded as an {\em effective} ``dark energy''
 component, which compiles the contributions to the Friedmann
 equation from the extra dimensions~\cite{r28,r29}. From
 Eq.~(\ref{eq4}), we have
 \be{eq6}
 1-\Omega_k=\Omega_m+\Omega_{de},
 \ee
 where $\Omega_k\equiv -k/(a^2H^2)$, and
 $\Omega_i\equiv (8\pi G\rho_i)/(3H^2)$ for $i=m$ and $de$. From
 Eqs.~(\ref{eq5}), (\ref{eq6}) and the energy conservation equation
 $\dot{\rho}_{de}+3H(\rho_{de}+p_{de})=0$, we find that the
 {\em effective} equation-of-state parameter~(EoS) for the
 {\em effective} ``dark energy'' component
 $w_{de}\equiv p_{de}/\rho_{de}$ is given by
 \be{eq7}
 w_{de}=\frac{-1+\Omega_k}{1+\Omega_m-\Omega_k}.
 \ee
 For the flat DGP model whose $\Omega_k=0$ exactly, we get
 $w_{de}=-(1+\Omega_m)^{-1}$, which agrees with the one
 of~\cite{r28,r29}. From Eq.~(\ref{eq3}), it is easy to see that
 \be{eq8}
 E^2(z)\equiv\left(\frac{H}{H_0}\right)^2=
 \left[\sqrt{\Omega_{m0}(1+z)^3+\Omega_{r_c}}
 +\sqrt{\Omega_{r_c}}\right]^2+\Omega_{k0}(1+z)^2,
 \ee
 where we have used $\rho_m=\rho_{m0}a^{-3}$; the subscript ``0''
 indicates the present value of the corresponding quantity;
 $\Omega_{r_c}\equiv (8\pi G\rho_{r_c})/(3H_0^2)=1/(4H_0^2 r_c^2)$.
 From Eq.~(\ref{eq8}), we have
 \be{eq9}
 1=\left[\sqrt{\Omega_{m0}+\Omega_{r_c}}
 +\sqrt{\Omega_{r_c}}\right]^2+\Omega_{k0}.
 \ee
 Therefore, only two of $\Omega_{m0}$, $\Omega_{r_c}$ and
 $\Omega_{k0}$ are independent model parameters. For the flat DGP
 model whose $\Omega_{k0}=0$ exactly, Eq.~(\ref{eq9}) becomes
 $\Omega_{m0}=1-2\sqrt{\Omega_{r_c}}$.

In Einstein gravity, the growth function $\delta(z)$ at scales much
 smaller than the Hubble radius obeys the following differential
 equation~\cite{r18,r19,r20,r21,r31,r40}
 \be{eq10}
 \ddot{\delta}+2H\dot{\delta}=4\pi G\rho_m\delta.
 \ee
 In modified gravity, Eq.~(\ref{eq10}) has been modified
 to~\cite{r27,r28,r29,r40}
 \be{eq11}
 \ddot{\delta}+2H\dot{\delta}=4\pi G_{\rm eff\,}\rho_m\delta,
 \ee
 where $G_{\rm eff}$ is the effective local gravitational
 ``constant'' measured by Cavendish-type experiment, which is
 time-dependent. In general, $G_{\rm eff}$ can be written as
 \be{eq12}
 G_{\rm eff}=G\left(1+\frac{1}{3\beta}\right),
 \ee
 where $\beta$ is determined once we specify the modified
 gravity theory. In the DGP gravity, $\beta$ is given
 by~\cite{r27,r29,r39,r40}
 \be{eq13}
 \beta=1-2r_cH\left(1+\frac{\dot{H}}{3H^2}\right).
 \ee
 By using Eq.~(\ref{eq3}), $\beta$ can be written as~\cite{r40}
 \be{eq14}
 1+\frac{1}{3\beta}=\frac{4\Omega_m^2-4\left(1-\Omega_k\right)^2
 +\alpha} {3\Omega_m^2-3\left(1-\Omega_k\right)^2+\alpha}\,,
 \ee
 where
 \be{eq15}
 \alpha\equiv 2\sqrt{1-\Omega_k}\left(3-4\Omega_k+2\Omega_m\Omega_k
 +\Omega_k^2\right).
 \ee
 For the flat DGP model whose $\Omega_k=0$ exactly, we get
 $\beta=-(1+\Omega_m^2)/(1-\Omega_m^2)$, which agrees with the
 one of~\cite{r28,r29}. Following~\cite{r21,r31}, we rewrite
 Eq.~(\ref{eq11}) as
 \be{eq16}
 \left(\ln\delta\right)^{\prime\prime}
 +\left(\ln\delta\right)^{\prime 2}
 +\left(2+\frac{H^\prime}{H}\right)\left(\ln\delta\right)^{\prime}
 =\frac{3}{2}\left(1+\frac{1}{3\beta}\right)\Omega_m,
 \ee
 where a prime denotes the derivative with respect to $\ln a$.
 By using Eqs.~(\ref{eq4}), (\ref{eq6}) and the energy conservation
 equation $\dot{\rho}_{de}+3H(1+w_{de})\rho_{de}=0$, we find that
 \be{eq17}
 \frac{H^\prime}{H}=\frac{\dot{H}}{H^2}=-\frac{3}{2}
 +\frac{\Omega_k}{2}-\frac{3}{2}w_{de}\left(1-\Omega_k
 -\Omega_m\right),
 \ee
 where $w_{de}$ is given in Eq.~(\ref{eq7}). Substituting
 Eq.~(\ref{eq17}) into Eq.~(\ref{eq16}), we obtain
 \be{eq18}
 \left(\ln\delta\right)^{\prime\prime}
 +\left(\ln\delta\right)^{\prime 2}
 +\left(\ln\delta\right)^{\prime}\left[\frac{1}{2}\left(1+
 \Omega_k\right)-\frac{3}{2}w_{de}\left(1-\Omega_k
 -\Omega_m\right)\right]
 =\frac{3}{2}\left(1+\frac{1}{3\beta}\right)\Omega_m,
 \ee
 In fact, the growth rate
 $f\equiv d\ln\delta/d\ln a=\left(\ln\delta\right)^{\prime}$. By
 using the definition of $\Omega_m$, the energy conservation
 equation $\dot{\rho}_m+3H\rho_m=0$ and Eq.~(\ref{eq17}), we have
 \be{eq19}
 \Omega_m^\prime=\Omega_m\left[3w_{de}\left(1-\Omega_k
 -\Omega_m\right)-\Omega_k\right].
 \ee
 Therefore, we find that
 \be{eq20}
 \left(\ln\delta\right)^{\prime\prime}=f^\prime=
 \Omega_m\left[3w_{de}\left(1-\Omega_k-\Omega_m\right)
 -\Omega_k\right]\frac{d\,f}{d\,\Omega_m}.
 \ee
 So, Eq.~(\ref{eq18}) becomes
 \be{eq21}
 \Omega_m\left[3w_{de}\left(1-\Omega_k-\Omega_m\right)
 -\Omega_k\right]\frac{d\,f}{d\,\Omega_m}+f^2
 +f\left[\frac{1}{2}\left(1+
 \Omega_k\right)-\frac{3}{2}w_{de}\left(1-\Omega_k
 -\Omega_m\right)\right]
 =\frac{3}{2}\left(1+\frac{1}{3\beta}\right)\Omega_m,
 \ee
 where $w_{de}$ and $\beta$ are given in Eqs.~(\ref{eq7})
 and~(\ref{eq14}), respectively. Substituting Eq.~(\ref{eq1}) into
 Eq.~(\ref{eq21}) and expanding around $\Omega_m=1$ (good
 approximation especially at $z\,\gsim\,1$), after some tedious
 algebra, we finally arrive at
 \be{eq22}
 10\gamma\left(\Omega_m-1\right)-\Omega_k\left(1-2\gamma\right)
 +3\left(1-2\gamma\right)\left(1-\Omega_m\right)
 =6\left(\Omega_m-1\right)+\Omega_m\left[\left(\Omega_m
 -1\right)\left(\Omega_m+1\right)+2\Omega_k\right],
 \ee
 where we have ignored the higher order terms of small quantities
 $1-\Omega_m$ and $\Omega_k$. Noting that Eq.~(\ref{eq6}), namely
 $1-\Omega_m=\Omega_k+\Omega_{de}$,
 we consider three cases for Eq.~(\ref{eq22}).

 \begin{description}
  \item[Case~(I)]
  $\Omega_k=0$ exactly or
  $\Omega_k\ll\Omega_{de}=1-\Omega_m\ll 1$.\\
  In this case, throwing out the terms of $\Omega_k$ in
  Eq.~(\ref{eq22}), then eliminating $\left(\Omega_m-1\right)$ in
  both sides, and using $\Omega_m\to 1$ finally, we have
  $\gamma=11/16$. Obviously, it agrees with the known one of the
  flat DGP model~\cite{r18}.

  \item[Case~(II)]
  $\Omega_{de}\ll\Omega_k=1-\Omega_m\ll 1$.\\
  In this case, eliminating $\left(\Omega_m-1\right)=-\Omega_k$ in
  both sides of Eq.~(\ref{eq22}) and then using $\Omega_m\to 1$,
  we have $\gamma=4/7$. In fact, it coincides with the curvature
  solution found firstly in~\cite{r51} and could be considered as
  a special case of the results in~\cite{r21,r18}  with $w=-1/3$.

  \item[Case~(III)]
  $\Omega_k\sim\Omega_{de}\sim 1-\Omega_m\ll 1$.\\
  In this case, $\Omega_k$, $\Omega_{de}$ and $1-\Omega_m$ are
  at the same order. Noting that $1-\Omega_m=\Omega_k+\Omega_{de}$,
  for convenience, we parameterize
  $\Omega_k=m\left(1-\Omega_m\right)$, where $0<m<1$. It is worth
  noting that generally $m$ is time-dependent and one considers
  only an instantaneous value of $m$. Substituting into
  Eq.~(\ref{eq22}), then eliminating $\left(\Omega_m-1\right)$
  in both sides, and using $\Omega_m\to 1$ finally, we have
  \be{eq23}
  \gamma=\frac{11-3m}{16-2m}.
  \ee
  Obviously, when $m\to 0$ and $1$, $\gamma\to 11/16$ and
  $4/7$, respectively.
 \end{description}

\noindent For the flat DGP model whose $\Omega_k=0$ always, the
 only theoretical growth index is given by $11/16$. For the DGP
 models whose $\Omega_k\not=0$, the situations are different.
 Noting that $\Omega_m\propto (1+z)^3$, $\Omega_k\propto (1+z)^2$
 and $\Omega_{de}\propto (1+z)^{3(1+w_{de})}$ with $w_{de}<-1/3$
 to accelerate the expansion of our universe, $\Omega_k$ increases
 faster than $\Omega_{de}$ when $z$ increases. So, for high $z$,
 $\Omega_{de}\ll\Omega_k=1-\Omega_m\ll 1$ eventually. Thus, we have
 $\gamma=4/7$ eventually, regardless of the value of $\Omega_{k0}$.
 However, if $\Omega_{k0}$ deviates from $0$ very small, it is still
 possible that $\Omega_k\ll\Omega_{de}=1-\Omega_m\ll 1$ at high $z$,
 where $\gamma=11/16$; eventually $\gamma=4/7$ at higher $z$. On the
 other hand, if $\Omega_{k0}$ deviates from $0$ not so small, the
 only theoretical growth index is given by $4/7$.

Finally, it is worth noting that the above results are obtained as
 high redshift asymptotic values, rather than values for the whole
 growth history or the asymptotic future. On the other hand, the
 general solution of $\gamma$ is obtained in~\cite{r18} for the
 flat DGP model. We refer to~\cite{r18} for details on this.

%============================= section 3 ===================================
\section{Confronting with current growth rate data}\label{sec3}
The most useful currently available growth rate data involve the
 redshift distortion parameter $\beta_L$~\cite{r41} observed through
 the anisotropic pattern of galactic redshifts on cluster
 scales~\cite{r42} (see also~\cite{r31}). The parameter $\beta_L$
 is related to the growth rate $f$ as~\cite{r31}
 \be{eq24}
 \beta_L=\frac{f}{b},
 \ee
 where $b$ is the bias factor. We present the currently available
 data for $\beta_L$ and $b$ at various redshifts in Table~\ref{tab1},
 along with the inferred growth rates. This is an extended version
 of the dataset used in Ref.~\cite{r31}, and contains a new data
 point at $z=0.77$~\cite{r45}. Hereafter, we call the five data
 points except the one at $z=0.77$ as dataset ``Fobs'' and call
 all the six data points as dataset ``Fobsext''.

About the currently available growth rate data, it is worth noting
 that there is considerable variation in analysis of different
 references and hence there is no accomplished consensus in fact.
 On the other hand, many of these results are not of the growth
 rate in isolation but fold in other large scale structure
 information. Besides, in Table~\ref{tab1}, some of the quoted
 values include the other datasets in the analysis, e.g. \cite{r46}
 includes \cite{r43} to derive their value of $\beta_L$. So, one
 should use these data with caution.

Using the growth rate data in Table~\ref{tab1}, we can perform
 a $\chi^2$ analysis to find the growth index $\gamma$ and
 check its consistency with the theoretical values. As well-known,
 the corresponding $\chi^2$ reads
 \be{eq25}
 \chi^2({\bf p},\gamma)=\sum\limits_i\left[\frac{f_{obs}(z_i)-
 f_{th}(z_i;{\bf p},\gamma)}{\sigma_{f_{obs}}}\right]^2,
 \ee
 where $f_{obs}$ and its corresponding $1\sigma$ uncertainty
 $\sigma_{f_{obs}}$ are given in Table~\ref{tab1}; $\bf p$ denotes
 the model parameters; $f_{th}(z_i;{\bf p},\gamma)$ can be obtained
 from Eq.~(\ref{eq1}), in which $\Omega_m$ can be rewritten as
 a more convenient form
 \be{eq26}
 \Omega_m(z)=\frac{\Omega_{m0}(1+z)^3}{E^2(z)},
 \ee
 and $E(z)$ can be found in Eq.~(\ref{eq8}).

%==================== table 1 ====================

 \begin{table}[tbp]
 \begin{center}
 \begin{tabular}{ccccc} \hline\hline
 $z$ & $\beta_L$ & $b$ & $f_{obs}$ & Reference\\ \hline
 $0.15$ & $0.49\pm 0.09$ & $1.04\pm 0.11$
 & $0.51\pm 0.11$ & \cite{r43}\\
 $0.35$ & $0.31\pm 0.04$ & $2.25\pm 0.08$
 & $0.70\pm 0.18$ & \cite{r7}\\
 $0.55$ & $0.45\pm 0.05$ & $1.66\pm 0.35$
 & $0.75\pm 0.18$ & \cite{r44}\\
 $0.77$ & $0.70\pm 0.26$ & $1.3\pm 0.1$
 & $0.91\pm 0.36$ & \cite{r45}\\
 $1.4$ & $0.60^{+0.14}_{-0.11}$ & $1.5\pm 0.20$
 & $0.90\pm 0.24$ & \cite{r46}\\
 $3.0$ & {\bf ---} & {\bf ---} & $1.46\pm 0.29$ & \cite{r47}\\
 \hline\hline
 \end{tabular}
 \end{center}
 \caption{\label{tab1} The currently available data for $\beta_L$
 and $b$ at various redshifts, along with the inferred growth
 rates. Notice that Ref.~\cite{r47} only reports the growth rate
 and not the $\beta_L$ and $b$ parameters, since the growth rate
 was obtained directly from the change of power spectrum
 $Ly-\alpha$ forest data in SDSS at various redshift slices. This
 is an extended version of the dataset used in Ref.~\cite{r31},
 and contains a new data point at $z=0.77$~\cite{r45}.}
 \end{table}

%=================================================

It is worth noting that the references in Table~\ref{tab1} have
 assumed flat $\Lambda$CDM (with $\Omega_{m0,\Lambda}=0.30$ for the
 five data points except the one at $z=0.77$, and with
 $\Omega_{m0,\Lambda}=0.25$ for the data point at $z=0.77$) when
 converting redshifts to distances for the power spectra and
 therefore their use to test models different from $\Lambda$CDM
 might be unreliable, as stressed in~\cite{r31}. However, we can get
 around this problem in a new way. As mentioned above, the key is
 the redshift-distance relation. If the DGP model and $\Lambda$CDM
 model share the same redshift-distance relation, these growth rate
 data can also be used in the DGP model. To this end, we should
 properly select the model parameters of DGP model in order to
 reproduce the same redshift-distance relation of $\Lambda$CDM
 model. There is a simple and efficient method. As well-known
 in any textbook, the comoving distance is given by
 \be{eq27}
 r(z)\equiv\frac{1}{H_0\sqrt{\left|\Omega_k\right|}}\,
 F\left(\sqrt{\left|\Omega_k\right|}\int_0^z
 \frac{d\tilde{z}}{E(\tilde{z})}\right),
 \ee
 where the function $F(x)=x$, $\sin x$ and $\sinh x$ for
 $\Omega_k=0$, $\Omega_k<0$ and $\Omega_k>0$, respectively.
 For convenience, we use the dimensionless comoving distance
 $H_0r(z)$ instead. It is easy to get the $H_0r(z)$ line for the
 flat $\Lambda$CDM model with $\Omega_{m0,\Lambda}=0.30$. Then, we
 discretize it into many points in a redshift range (for instance,
 $0\leq z\leq 4$, which covers the range of current growth data),
 and manually assign a relative ``error'' (say, $0.5\%$) to these
 discrete points. So, we have many {\em fake} ``data points'' in
 hand. Then, we fit the DGP model to these {\em fake} ``data
 points'' and find out the ``best fit'' parameters. Obviously,
 the DGP model with these ``best fit'' parameters will share the
 same redshift-distance relation with the flat $\Lambda$CDM model.
 Hence, the growth rate data in Table~\ref{tab1} can also be used
 for the corresponding DGP model. Finally, there is a minor remark
 on the procedure mentioned above. In fact, it holds only for the
 low redshift quantities considered here. Comparison of the
 distances to CMB last scattering ($z\sim 1100$) will show large
 differences between these matching models. Fortunately, the
 redshift of current growth data is less than $4$ and the procedure
 mentioned above works well.

%============================= Fig. 1 =================================

 \begin{center}
 \begin{figure}[htbp]
 \centering
 \includegraphics[width=0.98\textwidth]{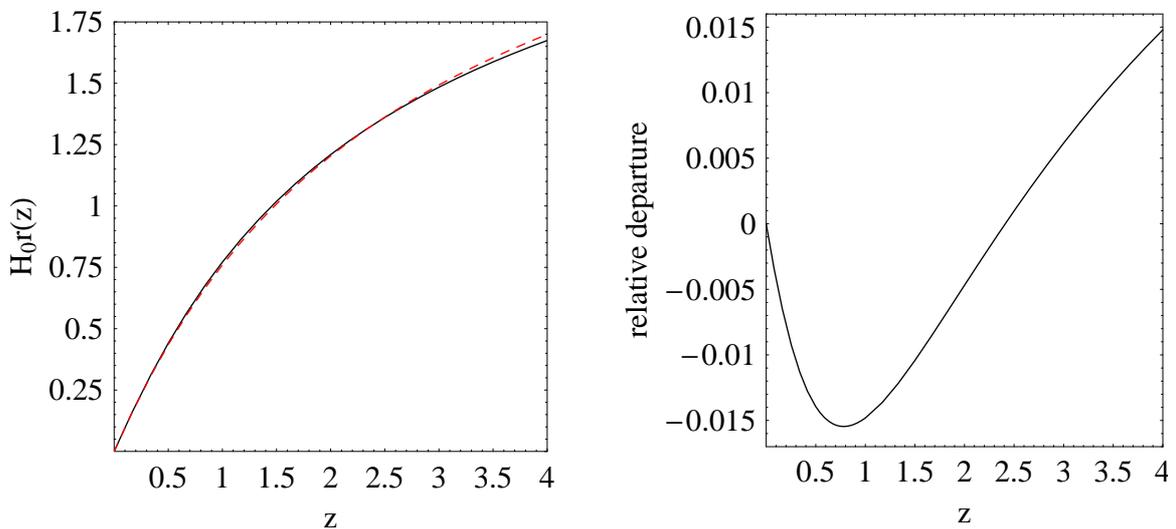}
 \caption{\label{fig1} The $H_0r(z)$ of flat DGP model with the
 corresponding ``best fit'' parameters (red dashed line) and flat
 $\Lambda$CDM model with $\Omega_{m0,\Lambda}=0.30$ (black solid
 line) are shown in left panel. The right panel shows the relative
 departure between these two $H_0r(z)$ lines.}
 \end{figure}
 \end{center}

%======================================================================

\begin{center}
 \begin{figure}[htbp]
 \centering
 \includegraphics[width=0.98\textwidth]{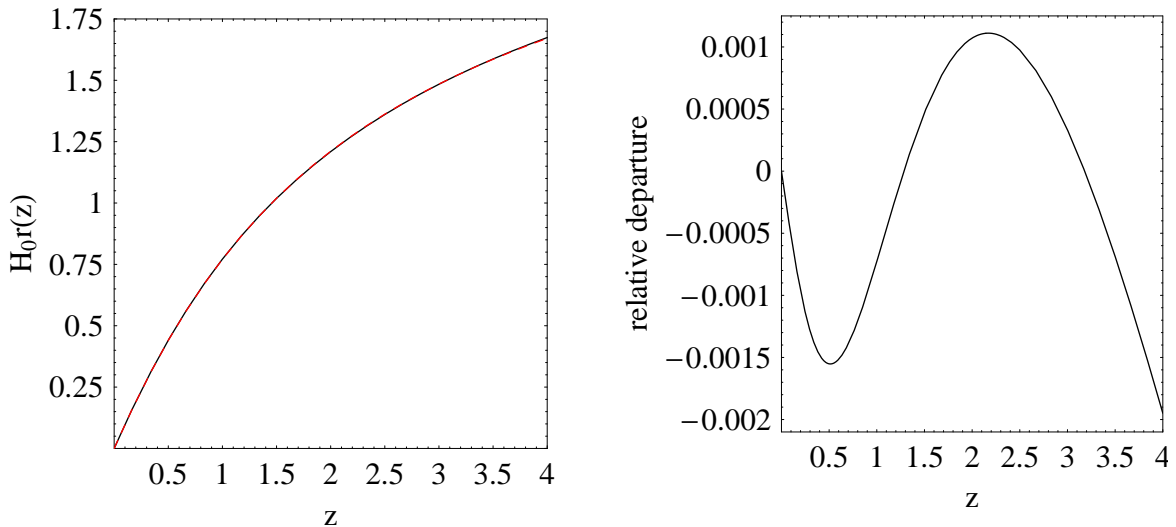}
 \caption{\label{fig2} The same as in Fig.~\ref{fig1},
 except for the DGP model with $\Omega_{k0}\not=0$.}
 \end{figure}
 \end{center}

We consider the flat DGP model whose $\Omega_{k0}=0$ exactly
 at first. Following above method, we find that the ``best fit''
 parameter of flat DGP model is $\Omega_{r_c}=0.1519$ (the
 corresponding $\Omega_{m0}=0.2204$). In Fig.~\ref{fig1}, we show
 the $H_0r(z)$ of the corresponding flat DGP model along with the
 one of flat $\Lambda$CDM model with $\Omega_{m0,\Lambda}=0.30$.
 The largest relative departure between these two $H_0r(z)$ lines
 in range $0\leq z\leq 4$ is less than $1.5\%$. Similarly, we
 consider the general DGP model with any $\Omega_{k0}$ and find
 that the ``best fit'' parameters of DGP model are
 $\Omega_{m0}=0.2217$ and $\Omega_{r_c}=0.1734$ (the corresponding
 $\Omega_{k0}=-0.0921$). In Fig.~\ref{fig2}, we show the $H_0r(z)$
 of the corresponding DGP model along with the one of flat
 $\Lambda$CDM model with $\Omega_{m0,\Lambda}=0.30$. Obviously,
 these two $H_0r(z)$ lines are degenerate in fact. The largest
 relative departure between the $H_0r(z)$ of DGP model with the
 corresponding ``best fit'' parameters and flat $\Lambda$CDM model
 with $\Omega_{m0,\Lambda}=0.30$ in range $0\leq z\leq 4$ is less
 than $0.19\%$ surprisingly. The DGP model with corresponding
 ``best fit'' parameters excellently reproduces the same
 redshift-distance relation of the flat $\Lambda$CDM model with
 $\Omega_{m0,\Lambda}=0.30$. So, the growth rate data in
 Table~\ref{tab1} can also be used for the corresponding DGP model.
 It is worth noting that all the above ``best fit'' parameters are
 consistent with the constraints of~\cite{r29,r48,r39} (see also
 e.g.~\cite{r34,r50}) and therefore they are also observationally
 acceptable.

%============================= Fig. 2 =================================

%======================================================================

Here, we fit the growth index $\gamma$ of these ``reproduced'' DGP
 models to the growth rate data in Table~\ref{tab1}. For the case
 of flat DGP model whose $\Omega_{k0}=0$ exactly, setting
 $\Omega_{m0}$ and $\Omega_{r_c}$ as the corresponding ``best fit''
 values and minimizing $\chi^2$ with respect to $\gamma$, we find
 that for dataset Fobs,
 $\gamma=0.438^{+0.126}_{-0.111}$ with $1\sigma$ uncertainty, or
 $\gamma=0.438^{+0.272}_{-0.209}$ with $2\sigma$ uncertainty; for
 dataset Fobsext, $\gamma=0.429^{+0.123}_{-0.108}$ with $1\sigma$
 uncertainty, or $\gamma=0.429^{+0.264}_{-0.205}$ with $2\sigma$
 uncertainty. Obviously, the only theoretical growth index
 $\gamma=11/16=0.6875$ is consistent with these results at
 $2\sigma$ uncertainty.

For the case of DGP model with $\Omega_{k0}\not=0$, setting
 $\Omega_{m0}$, $\Omega_{r_c}$ and $\Omega_{k0}$ as the
 corresponding ``best fit'' values and minimizing $\chi^2$ with
 respect to $\gamma$, we find that for dataset Fobs,
 $\gamma=0.465^{+0.134}_{-0.117}$ with $1\sigma$ uncertainty, or
 $\gamma=0.465^{+0.290}_{-0.221}$ with $2\sigma$ uncertainty; for
 dataset Fobsext, $\gamma=0.457^{+0.131}_{-0.115}$ with $1\sigma$
 uncertainty, or $\gamma=0.457^{+0.282}_{-0.217}$ with $2\sigma$
 uncertainty. Obviously, the theoretical growth index
 $\gamma=11/16=0.6875$ is consistent with these results at
 $2\sigma$ uncertainty, whereas the other theoretical growth index
 $\gamma=4/7\simeq 0.5714$ is consistent with these results at
 $1\sigma$ uncertainty.

%============================= section 4 ===================================
\section{Concluding remarks}\label{sec4}
In summary, we consider the growth index of general DGP model with
 any $\Omega_k$ in this note. We confront the growth index of DGP
 model with current growth rate data and find that the DGP model
 is still consistent with it. On the other hand, it is shown that
 the $\Lambda$CDM model is also consistent with current growth rate
 data~\cite{r31}. This implies that more and better growth rate
 data are required to distinguish between dark energy and modified
 gravity models.

Some remarks are in order. Firstly, it is worth noting that the
 references in Table~\ref{tab1} have assumed flat $\Lambda$CDM
 (with $\Omega_{m0,\Lambda}=0.30$ for the five data points except
 the one at $z=0.77$, and with $\Omega_{m0,\Lambda}=0.25$ for the
 data point at $z=0.77$) when converting redshifts to distances
 for the power spectra and therefore their use to test models
 different from $\Lambda$CDM might be unreliable, as stressed
 in~\cite{r31}. Although we can get around this problem by means
 of reproducing the same redshift-distance relation between
 the DGP model and $\Lambda$CDM model, the really complete solution
 to this problem lies on the reanalyzing the power spectra with the
 proper redshift-distance relation in the corresponding modified
 gravity. Secondly, in fact, we fixed $\Omega_{m0}$ while fitting
 for $\gamma$ in section~\ref{sec3}. This is required to ensure
 the validity of matching procedure of finding an equivalent DGP
 model. We admit that this weakens our conclusion. It is desirable
 to find a new method to improve the validity of matching
 procedure in the future works. Thirdly, one may use other
 parameterizations to the growth rate $f$, such as
 $f=\Omega_m^\gamma(1+\eta)$~\cite{r30}, or $f=\Omega_m^\gamma$
 with $\gamma=\gamma_0+\gamma_0^\prime z$~\cite{r38}. However, the
 additional parameters $\eta$ and $\gamma_0^\prime$ are found to
 be negligible in fact~\cite{r30,r31,r38}. On the other hand, since
 currently available growth rate data points are so few and have
 large errors, it is difficult to tightly constrain these
 additional parameters. Fourthly, although as shown in~\cite{r23}
 that non-trivial dark energy clustering or interaction between
 dark energy and dark matter might bring some troubles,
 we consider that combining the probes of expansion history and
 growth history is still promising to distinguish the dark energy
 and modified gravity. Of course, new idea to distinguish dark
 energy and modified gravity is still desirable.

%============================= Acknowledgments ===================================

\section*{Acknowledgments}
We thank the anonymous referee for expert comments and quite useful
 suggestions, which help us to improve this work. We are
 grateful to Professor Shuang~Nan~Zhang and Professor Rong-Gen~Cai
 for helpful discussions. We also thank Minzi~Feng,
 as well as Hong-Sheng~Zhang, Nan~Liang, Rong-Jia~Yang, Wei-Ke~Xiao,
 Pu-Xun~Wu, Jian~Wang, Lin~Lin and Yuan~Liu, for kind
 help and discussions. The initial part of this work was completed
 during the period of Chinese New Year 2008, with an impressive
 disaster of snowstorm and cold rain in southern China. We
 acknowledge partial funding support from China Postdoctoral Science
 Foundation, and by the Ministry of Education of China, Directional
 Research Project of the Chinese Academy of Sciences under project
 No.~KJCX2-YW-T03, and by the National Natural Science Foundation
 of China under project No.~10521001.

%============================= References ====================================


\begin{thebibliography}{99}

\bibitem{r1}
P.~J.~E.~Peebles and B.~Ratra,
 Rev.\ Mod.\ Phys.\  {\bf 75}, 559 (2003) [astro-ph/0207347];\\
T.~Padmanabhan, Phys.\ Rept.\  {\bf 380}, 235 (2003) [hep-th/0212290];\\
S.~M.~Carroll, astro-ph/0310342;\\
R.~Bean, S.~Carroll and M.~Trodden, astro-ph/0510059;\\
V.~Sahni and A.~A.~Starobinsky,
 Int.\ J.\ Mod.\ Phys.\ D {\bf 9}, 373 (2000) [astro-ph/9904398];\\
S.~M.~Carroll, Living Rev.\ Rel.\  {\bf 4}, 1 (2001) [astro-ph/0004075];\\
T.~Padmanabhan, Curr.\ Sci.\  {\bf 88}, 1057 (2005) [astro-ph/0411044];\\
S.~Weinberg, Rev.\ Mod.\ Phys.\  {\bf 61}, 1 (1989);\\
S.~Nobbenhuis,
 Found.\ Phys.\  {\bf 36}, 613 (2006) [gr-qc/0411093];\\
E.~J.~Copeland, M.~Sami and S.~Tsujikawa,
 Int.\ J.\ Mod.\ Phys.\  D {\bf 15}, 1753 (2006)
 [hep-th/0603057];\\
A.~Albrecht {\it et al.}, astro-ph/0609591;\\
R.~Trotta and R.~Bower, astro-ph/0607066;\\
M.~Kamionkowski, arXiv:0706.2986 [astro-ph];\\
B.~Ratra and M.~S.~Vogeley, arXiv:0706.1565 [astro-ph];\\
E.~V.~Linder, arXiv:0705.4102 [astro-ph];\\
M.~S.~Turner and D.~Huterer, arXiv:0706.2186 [astro-ph];\\
J.~Frieman, M.~Turner and D.~Huterer, arXiv:0803.0982 [astro-ph].

\bibitem{r2}
A.~G.~Riess {\it et al.} [Supernova Search Team Collaboration],
 Astron.\ J.\  {\bf 116}, 1009 (1998) [astro-ph/9805201];\\
S.~Perlmutter {\it et al.} [Supernova Cosmology Project
Collaboration], Astrophys.\ J.\  {\bf 517}, 565 (1999)
[astro-ph/9812133];\\
J.~L.~Tonry {\it et al.}  [Supernova Search Team Collaboration],
 Astrophys.\ J.\  {\bf 594}, 1 (2003) [astro-ph/0305008];\\
R.~A.~Knop {\it et al.} [Supernova Cosmology Project
 Collaboration], Astrophys.\ J.\  {\bf 598}, 102 (2003)
 [astro-ph/0309368];\\
A.~G.~Riess {\it et al.} [Supernova Search Team Collaboration],
 Astrophys.\ J.\  {\bf 607}, 665 (2004) [astro-ph/0402512].

\bibitem{r3}
A.~G.~Riess {\it et al.} [Supernova Search Team Collaboration],
 Astrophys.\ J.\  {\bf 659}, 98 (2007) [astro-ph/0611572].

\bibitem{r4}
P.~Astier {\it et al.} [SNLS Collaboration],
 Astron.\ Astrophys.\  {\bf 447}, 31 (2006) [astro-ph/0510447];\\
J.~D.~Neill {\it et al.},
 Astron.\ J.\  {\bf 132}, 1126 (2006) [astro-ph/0605148].

\bibitem{r5}
D.~N.~Spergel {\it et al.}  [WMAP Collaboration],
 Astrophys.\ J.\ Suppl.\  {\bf 170}, 377 (2007)
 [astro-ph/0603449];\\
L.~Page {\it et al.}  [WMAP Collaboration],
 Astrophys.\ J.\ Suppl.\  {\bf 170}, 335 (2007)
 [astro-ph/0603450];\\
G.~Hinshaw {\it et al.}  [WMAP Collaboration],
 Astrophys.\ J.\ Suppl.\  {\bf 170}, 288 (2007)
 [astro-ph/0603451];\\
N.~Jarosik {\it et al.}  [WMAP Collaboration],
 Astrophys.\ J.\ Suppl.\  {\bf 170}, 263 (2007)
 [astro-ph/0603452].

\bibitem{r6}
M.~Tegmark {\it et al.} [SDSS Collaboration],
 Phys.\ Rev.\ D {\bf 69}, 103501 (2004) [astro-ph/0310723];\\
M.~Tegmark {\it et al.} [SDSS Collaboration],
 Astrophys.\ J.\  {\bf 606}, 702 (2004) [astro-ph/0310725];\\
U.~Seljak {\it et al.},
 Phys.\ Rev.\ D {\bf 71}, 103515 (2005) [astro-ph/0407372];\\
J.~K.~Adelman-McCarthy {\it et al.}  [SDSS Collaboration],
 Astrophys.\ J.\ Suppl.\  {\bf 162}, 38 (2006) [astro-ph/0507711];\\
K.~Abazajian {\it et al.} [SDSS Collaboration], astro-ph/0410239;
 astro-ph/0403325; astro-ph/0305492.

\bibitem{r7}
M.~Tegmark {\it et al.} [SDSS Collaboration],
 Phys.\ Rev.\  D {\bf 74}, 123507 (2006) [astro-ph/0608632].

\bibitem{r8}
S.~W.~Allen {\it et al.},
 Mon.\ Not.\ Roy.\ Astron.\ Soc.\  {\bf 353}, 457 (2004)
 [astro-ph/0405340];\\
S.~W.~Allen {\it et al.}, arXiv:0706.0033 [astro-ph];\\
A.~Mantz, S.~W.~Allen, H.~Ebeling and D.~Rapetti,
 arXiv:0709.4294 [astro-ph].

\bibitem{r9}
W.~M.~Wood-Vasey {\it et al.}  [ESSENCE Collaboration],
 Astrophys.\ J.\  {\bf 666}, 694 (2007) [astro-ph/0701041];\\
G.~Miknaitis {\it et al.} [ESSENCE Collaboration],
  Astrophys.\ J.\  {\bf 666}, 674 (2007)
  [astro-ph/0701043].

\bibitem{r10}
C.~Shapiro and M.~S.~Turner,
 Astrophys.\ J.\  {\bf 649}, 563 (2006) [astro-ph/0512586].

\bibitem{r11}
M.~Seikel and D.~J.~Schwarz, arXiv:0711.3180 [astro-ph];\\
Y.~Gong, A.~Wang, Q.~Wu and Y.~Z.~Zhang,
 JCAP {\bf 0708}, 018 (2007) [astro-ph/0703583];\\
Y.~Gong and A.~Wang,
 Phys.\ Rev.\  D {\bf 73}, 083506 (2006) [astro-ph/0601453].

\bibitem{r12}
R.~R.~Caldwell and E.~V.~Linder,
 Phys.\ Rev.\ Lett.\  {\bf 95}, 141301 (2005)
 [astro-ph/0505494];\\
E.~V.~Linder,
 Phys.\ Rev.\  D {\bf 73}, 063010 (2006) [astro-ph/0601052].

\bibitem{r13}
R.~J.~Scherrer,
 Phys.\ Rev.\  D {\bf 73}, 043502 (2006) [astro-ph/0509890].

\bibitem{r14}
T.~Chiba,
 Phys.\ Rev.\  D {\bf 73}, 063501 (2006) [astro-ph/0510598].

\bibitem{r15}
E.~V.~Linder,
 Gen.\ Rel.\ Grav.\  {\bf 40}, 329 (2008) [arXiv:0704.2064].

\bibitem{r16}
V.~Sahni, T.~D.~Saini, A.~A.~Starobinsky and U.~Alam,
 JETP Lett.\  {\bf 77}, 201 (2003) [astro-ph/0201498];\\
U.~Alam, V.~Sahni, T.~D.~Saini and A.~A.~Starobinsky,
 Mon.\ Not.\ Roy.\ Astron.\ Soc.\  {\bf 344}, 1057 (2003)
 [astro-ph/0303009].

\bibitem{r17}
H.~Wei and R.~G.~Cai,
 Phys.\ Lett.\  B {\bf 655}, 1 (2007) [arXiv:0707.4526].

\bibitem{r18}
E.~V.~Linder and R.~N.~Cahn,
 Astropart.\ Phys.\  {\bf 28}, 481 (2007) [astro-ph/0701317].

\bibitem{r19}
E.~V.~Linder,
 Phys.\ Rev.\  D {\bf 72}, 043529 (2005) [astro-ph/0507263].

\bibitem{r20}
D.~Huterer and E.~V.~Linder,
 Phys.\ Rev.\  D {\bf 75}, 023519 (2007) [astro-ph/0608681].

\bibitem{r21}
L.~M.~Wang and P.~J.~Steinhardt,
 Astrophys.\ J.\  {\bf 508}, 483 (1998) [astro-ph/9804015].

\bibitem{r22}
Y.~Wang, arXiv:0710.3885 [astro-ph];\\
Y.~Wang, arXiv:0712.0041 [astro-ph].

\bibitem{r23}
M.~Kunz and D.~Sapone,
 Phys.\ Rev.\ Lett.\  {\bf 98}, 121301 (2007) [astro-ph/0612452];\\
E.~Bertschinger and P.~Zukin, arXiv:0801.2431 [astro-ph];\\
H.~Wei and S.~N.~Zhang, arXiv:0803.3292 [astro-ph], accepted
 for publication in Phys.\ Rev.\  D.

\bibitem{r24}
S.~Wang, L.~Hui, M.~May and Z.~Haiman,
 Phys.\ Rev.\  D {\bf 76}, 063503 (2007) [arXiv:0705.0165].

\bibitem{r25}
A.~Cardoso, K.~Koyama, S.~S.~Seahra and F.~P.~Silva,
 arXiv:0711.2563 [astro-ph].

\bibitem{r26}
K.~Koyama,
 Gen.\ Rel.\ Grav.\  {\bf 40}, 421 (2008) [arXiv:0706.1557].

\bibitem{r27}
K.~Koyama and R.~Maartens,
 JCAP {\bf 0601}, 016 (2006) [astro-ph/0511634].

\bibitem{r28}
A.~Lue, R.~Scoccimarro and G.~D.~Starkman,
 Phys.\ Rev.\  D {\bf 69}, 124015 (2004) [astro-ph/0401515].

\bibitem{r29}
A.~Lue,
 Phys.\ Rept.\  {\bf 423}, 1 (2006) [astro-ph/0510068].

\bibitem{r30}
C.~Di Porto and L.~Amendola, arXiv:0707.2686 [astro-ph];\\
L.~Amendola, M.~Kunz and D.~Sapone, arXiv:0704.2421 [astro-ph];\\
D.~Sapone and L.~Amendola, arXiv:0709.2792 [astro-ph].

\bibitem{r31}
S.~Nesseris and L.~Perivolaropoulos,
 Phys.\ Rev.\  D {\bf 77}, 023504 (2008) [arXiv:0710.1092].

\bibitem{r32}
G.~R.~Dvali, G.~Gabadadze and M.~Porrati,
 Phys.\ Lett.\  B {\bf 485}, 208 (2000) [hep-th/0005016].

\bibitem{r33}
C.~Deffayet,
 Phys.\ Lett.\  B {\bf 502}, 199 (2001) [hep-th/0010186];\\
C.~Deffayet, G.~R.~Dvali and G.~Gabadadze,
 Phys.\ Rev.\  D {\bf 65}, 044023 (2002) [astro-ph/0105068].

\bibitem{r34}
Z.~K.~Guo, Z.~H.~Zhu, J.~S.~Alcaniz and Y.~Z.~Zhang,
 Astrophys.\ J.\  {\bf 646}, 1 (2006) [astro-ph/0603632].

\bibitem{r35}
I.~Sawicki, Y.~S.~Song and W.~Hu,
 Phys.\ Rev.\  D {\bf 75}, 064002 (2007) [astro-ph/0606285].

\bibitem{r36}
P.~J.~E.~Peebles, {\it Large-Scale Structure of the Universe},
 Princeton University Press (1980);\\
P.~J.~E.~Peebles,
 Astrophys.\ J.\  {\bf 284}, 439 (1984).

\bibitem{r37}
O.~Lahav, P.~B.~Lilje, J.~R.~Primack and M.~J.~Rees,
 Mon.\ Not.\ Roy.\ Astron.\ Soc.\  {\bf 251}, 128 (1991).

\bibitem{r38}
D.~Polarski and R.~Gannouji,
 Phys.\ Lett.\  B {\bf 660}, 439 (2008) [arXiv:0710.1510];\\
R.~Gannouji and D.~Polarski, arXiv:0802.4196 [astro-ph].

\bibitem{r39}
M.~S.~Movahed, M.~Farhang and S.~Rahvar, astro-ph/0701339.

\bibitem{r40}
T.~Chiba and R.~Takahashi,
 Phys.\ Rev.\  D {\bf 75}, 101301 (2007) [astro-ph/0703347].

\bibitem{r41}
A.~J.~S.~Hamilton, astro-ph/9708102.

\bibitem{r42}
S.~Nesseris and L.~Perivolaropoulos,
 JCAP {\bf 0701}, 018 (2007) [astro-ph/0610092].

\bibitem{r43}
E.~Hawkins {\it et al.},
 Mon.\ Not.\ Roy.\ Astron.\ Soc.\  {\bf 346}, 78 (2003)
 [astro-ph/0212375];\\
L.~Verde {\it et al.},
 Mon.\ Not.\ Roy.\ Astron.\ Soc.\  {\bf 335}, 432 (2002)
 [astro-ph/0112161];\\
E.~V.~Linder, arXiv:0709.1113 [astro-ph].

\bibitem{r44}
N.~P.~Ross {\it et al.}, astro-ph/0612400.

\bibitem{r45}
L.~Guzzo {\it et al.}, Nature\ {\bf 451}, 541 (2008)
 [arXiv:0802.1944].

\bibitem{r46}
J.~da Angela {\it et al.}, astro-ph/0612401.

\bibitem{r47}
P.~McDonald {\it et al.}  [SDSS Collaboration],
 Astrophys.\ J.\  {\bf 635}, 761 (2005) [astro-ph/0407377].

\bibitem{r48}
J.~S.~Alcaniz and N.~Pires,
 Phys.\ Rev.\  D {\bf 70}, 047303 (2004) [astro-ph/0404146].

\bibitem{r49}
T.~M.~Davis {\it et al.},
 Astrophys.\ J.\  {\bf 666}, 716 (2007) [astro-ph/0701510].\\
 It compiled the joint 192 SNIa data from ESSENCE~\cite{r9}
 and new~Gold~\cite{r3}.\\
 The numerical data of the full sample are available at\\
 http:$/\!/$www.ctio.noao.edu/essence
 ~~~or~~~ http:$/\!/$braeburn.pha.jhu.edu/$^\sim$ariess/R06

\bibitem{r50}
V.~Barger, Y.~Gao and D.~Marfatia,
 Phys.\ Lett.\  B {\bf 648}, 127 (2007) [astro-ph/0611775];\\
B.~Wang, Y.~G.~Gong and R.~K.~Su,
 Phys.\ Lett.\  B {\bf 605}, 9 (2005) [hep-th/0408032].

\bibitem{r51}
J.~N.~Fry, Phys.\ Lett.\  B {\bf 158}, 211 (1985).

\end{thebibliography}
\end{document}